\documentclass[twocolumn,prc,showpacs,amsmath,amssymb,superscriptaddress]{revtex4}

\usepackage{graphicx}
\usepackage{dcolumn}
\usepackage{bm}

\begin{document}

\title{
Onset of intruder ground state in exotic Na isotopes
and evolution of the \mbox{\boldmath{$N=20$}} shell gap
}

\author{Yutaka Utsuno}
\affiliation{Japan Atomic Energy Research Institute, Tokai, Ibaraki
              319-1195, Japan}
\author{Takaharu Otsuka}
\affiliation{Department of Physics, University of Tokyo, Hongo, Tokyo
          113-0033, Japan}
\affiliation{Center for Nuclear Study, University of Tokyo, Hongo, Tokyo
          113-0033, Japan}
\affiliation{RIKEN, Hirosawa, Wako-shi, Saitama 351-0198, Japan}
\author{Thomas Glasmacher}
\affiliation{National Superconducting Cyclotron Laboratory and Department
of Physics and Astronomy, Michigan State University, East Lansing,
MI 48824, USA}
\author{Takahiro Mizusaki}
\affiliation{Institute of Natural Sciences,
             Senshu University, Higashimita, Tama,
              Kawasaki, Kanagawa 214-8580, Japan}
\author{Michio Honma}
\affiliation{Center for Mathematical Sciences, University of Aizu, Tsuruga,
             Ikki-machi Aizu-Wakamatsu, Fukushima 965-8580, Japan}
\date{\today}

\begin{abstract}

The onset of intruder ground states in Na isotopes is investigated
by comparing experimental data and shell-model calculations.
This onset is one of the consequences of the disappearance of the
$N$=20 magic structure, and the Na isotopes are shown to play a
special role in clarifying the change of this magic structure.
Both the electromagnetic moments and the energy levels
clearly indicate an onset of ground state intruder configurations at
neutron number $N=19$ already,
which arises only with a narrow $N=20$ shell gap in Na isotopes
resulting from the spin-isospin dependence of the
nucleon-nucleon interaction (as compared to a wider gap in
stable nuclei like $^{40}$Ca).  It is shown why the previous report
based on the mass led to a wrong conclusion.

\end{abstract}


\maketitle

\section{introduction}\label{sec:intro}

Amongst the most intriguing and unique features in exotic
nuclei are rather significant changes from the conventional magic structure.
As a result of them, the ground state of a nucleus with
$N$~or~$Z$ close to a conventional magic number
is not necessarily spherical, and can be strongly deformed.
Its fingerprint was first identified from
extra binding energies of $^{31,32}$Na \cite{na31mass},
whose origin was regarded, consistently with a Hartree-Fock calculation
\cite{campi}, as the dominance of strongly deformed intruder components
in the ground state over the normal components.
Here, normal (intruder) states imply the states comprised of shell-model
configurations without (with) $1p1h$, $2p2h$ or higher excited configurations
across the $N=20$ shell gap.
Later, more direct experimental evidence
of the strong deformation was found for $^{32}$Mg from the
low excitation energy of the $2^+_1$ state
\cite{mg32ex1,mg32ex2} and the large
$B(E2; 0^+_1 \to 2^+_1)$ \cite{mg32be2} value.
Thus, the disappearance of the $N$=20 magic
structure has been established in some $N=20$ isotones including
the recent case for $^{30}$Ne \cite{ne30}.
It still remains, however, an open question as to where the ground state
changes
from a normal- to an intruder-dominant configuration
in the chain of isotopes, and the question as to
what mechanism drives this disappearance remains.
The present paper aims at presenting the resolution of these
questions, as exemplified in the structure of Na isotopes.

For Na isotopes, one may expect
that the onset of the intruder-dominance
of the ground state lies right at $N=20$, from the
comparison of the experimental mass to a shell-model result
within the $sd$ shell with the USD interaction \cite{usd}.
The USD interaction has been the most frequently used interaction
in the $sd$ shell, and we shall refer to shell model calculations
with this interaction in the $sd$ shell as USD model or
calculation, hereafter.
A similar picture about the onset is assumed in the so-called
``island of inversion'' model \cite{island,caurier-n20},
where the lowest normal and the lowest intruder states are
confronted without mixing between them.
Although the mass (or the separation energy) often
provides us with helpful information on shell
structure, studies from different angles are needed before one draws
definite conclusions, as we shall demonstrate.
The first part of the present paper is focused upon
re-examination on the dominant configuration of the
ground state of Na isotopes.
We perform a large-scale shell-model calculation using the
Monte Carlo shell model (MCSM) \cite{mcsm-rev}, which is briefly
described in Sec.~\ref{sec:sm}.
In Sec.~\ref{sec:mom} and \ref{sec:level},  respectively,
the electromagnetic moments and the energy levels are
discussed, and from such discussions the transition point from the normal-
to intruder-dominant ground state is identified in the chain of
Na isotopes.
In Sec.~\ref{sec:gap}, the second part of the present paper, 
we discuss the mechanism of the
disappearance of the magic structure, focusing upon the
(effective) $N$=20 gap between the $sd$ and $pf$ shells and
emphasizing the special importance of the nucleus $^{30}$Na
on this issue from a somewhat general viewpoint.
We finally summarize the present study in Sec.~\ref{sec:summary}.

\section{Outline of the shell model calculation}\label{sec:sm}

The model space and the effective interaction used in the
present study are the same as those of our previous
studies \cite{mcsm-n20,mcsm-f}:
the valence shell consists of the full $sd$-shell orbits and
two lower $pf$-shell orbits.
The effective interaction is called hereafter
{\it SDPF-M} for the sake of clarification from other interactions.
The SDPF-M interaction was introduced in Ref.~\cite{mcsm-n20} in 1999,
by combining the USD interaction \cite{usd} for the $sd$ shell,
the Kuo-Brown interaction \cite{kb} for the $pf$ shell,
and a modified Millener-Kurath interaction \cite{mk-a40} for
the cross shell.
On top of this, a small but important modification was made for its
monopole part \cite{mcsm-n20} as we shall add some remarks later.
A unique feature of the SDPF-M interaction is that the neutron shell
structure, defined by the {\it effective single-particle energy}
(ESPE), changes, as a function of the proton number,
more significantly than in
previous models, for instance, the ``island of inversion''
\cite{island,caurier-n20}.  Here, the ESPE
includes mean effects from other valence nucleons on top of the
usual single-particle energies with respect to the given inert core
(i.e., closed shell).
Therefore, the ESPE depends on shell-model interactions between
valence nucleons.  The present varying shell structure can be explained
by the shell evolution mechanism of Ref. \cite{magic}
in terms of the spin-isospin property of the effective
nucleon-nucleon ($NN$) interaction.
The strong $T=0$ monopole attraction between the
$0d_{3/2}$ and the $0d_{5/2}$ enlarges the $N=20$ gap,
as protons occupy $0d_{5/2}$.
Inversely, this effect diminishes towards $Z=8$,
ending up with a rather narrow $N=20$ gap and a wider $N=16$ gap.
This shell evolution leads us to the oxygen drip line at $N=16$
\cite{o26ganil,o26msu,o28ganil,sakurai} as a result of emerging
$N=16$ magic number \cite{n16}.
The monopole part of the SDPF-M interaction was modified from that of
the USD interaction so as to reproduce the oxygen drip line \cite{mcsm-n20},
while the resultant monopole part is closer to the G-matrix result
as emphasized in \cite{magic}.

In the $N=20$ region, as we shall illustrate,
the Na isotopes give indispensable information on this shell evolution:
(i) specific Na isotopes provide us with clues of a narrow $N=20$ shell gap,
(ii) with odd $Z$, their ground-state properties can be
directly examined by non-vanishing electromagnetic moments,
and (iii) many experimental data have been recently accumulated
about the mass \cite{na-mass}, moment \cite{keim1,keim2},
$\gamma$-ray spectrum and transition by the
Coulomb excitation \cite{na31coul,na30coul}, etc.
Thus, we carry out shell-model studies on Na isotopes from
$N$=16 to 20.

Since the dimension of the Hamiltonian matrix
becomes prohibitively large
with the present problems,
we perform a shell-model calculation by the MCSM
based on the quantum Monte Carlo diagonalization (QMCD) method
whose development has been described in
\cite{qmcd-ph1,qmcd-proj,qmcd-ph2,qmcd-ph3}.
In the present MCSM calculation, we adopt the so-called
{\it $J$-compressed} bases \cite{qmcd-ph3}, i.e., bases
generated and adopted by monitoring the energy with the full
angular momentum projection.
The feasibility of the MCSM calculation for odd-$A$ nuclei
has been demonstrated in \cite{mcsm-f}.
This method works very well for odd-odd nuclei as well.

In the present calculation,
the $E2$ matrix elements are calculated with the
effective charges $(e_p, e_n)=(1.3e,0.5e)$ which are the
same as those used in the USD model \cite{usd}.
It has been confirmed that the MCSM with
these effective charges excellently
reproduce the $B(E2;0^+_1 \to 2^+_1)$ values of
even-even nuclei from stable to unstable nuclei
\cite{mcsm-n20}.
As for the effective $M1$ operator,
Brown and Wildenthal took an empirically optimum one
within the USD model \cite{g-sd}. They found that
the free-nucleon $g$ factors give
no obviously deviating magnetic moments but more quantitative
agreement can be attained
with the empirically optimum operator:
for $A=28$, $g_s$ is quenched by a factor 0.85, and
$g_l^{\rm p}=1.127$, $g_l^{\rm n}=-0.089$, $g_p^{\rm p}=0.041$ and
$g_p^{\rm n}=-0.35$ are used
where $l$, $s$, and $p$ are the orbital angular momentum,
the intrinsic spin, and
$\sqrt{8\pi}[Y^{(2)}(\mbox{\boldmath $r$})\otimes s]^{(1)}$
operators, respectively.
Recently, Honma {\it et al.} have presented in \cite{honma-pf}
that the spin $g$ factor does not have to be much quenched
in the full $pf$-shell model space
using their newly developed interaction \cite{gxpf1,honma-pf}.
Based on these extensive shell-model studies,
the $g$ factors are adopted, in the present MCSM calculations,
so as to be rather close to the above-mentioned
ones. Namely the spin part is quenched by a factor 0.9, and the other
$g$ factors
are shifted from the free-nucleon values by
$\delta g_l(\rm IV)=0.15$ and $\delta g_p(\rm IV)=0.5$
where $g({\rm IV})$ denotes the isovector
$g$ factor defined by
$g({\rm IV})$\,=\,$(g^{\rm p}-g^{\rm n})/2$.

\section{Electromagnetic moments}\label{sec:mom}

\begin{figure}
 \begin{center}
 \includegraphics[width=7.0cm,clip]{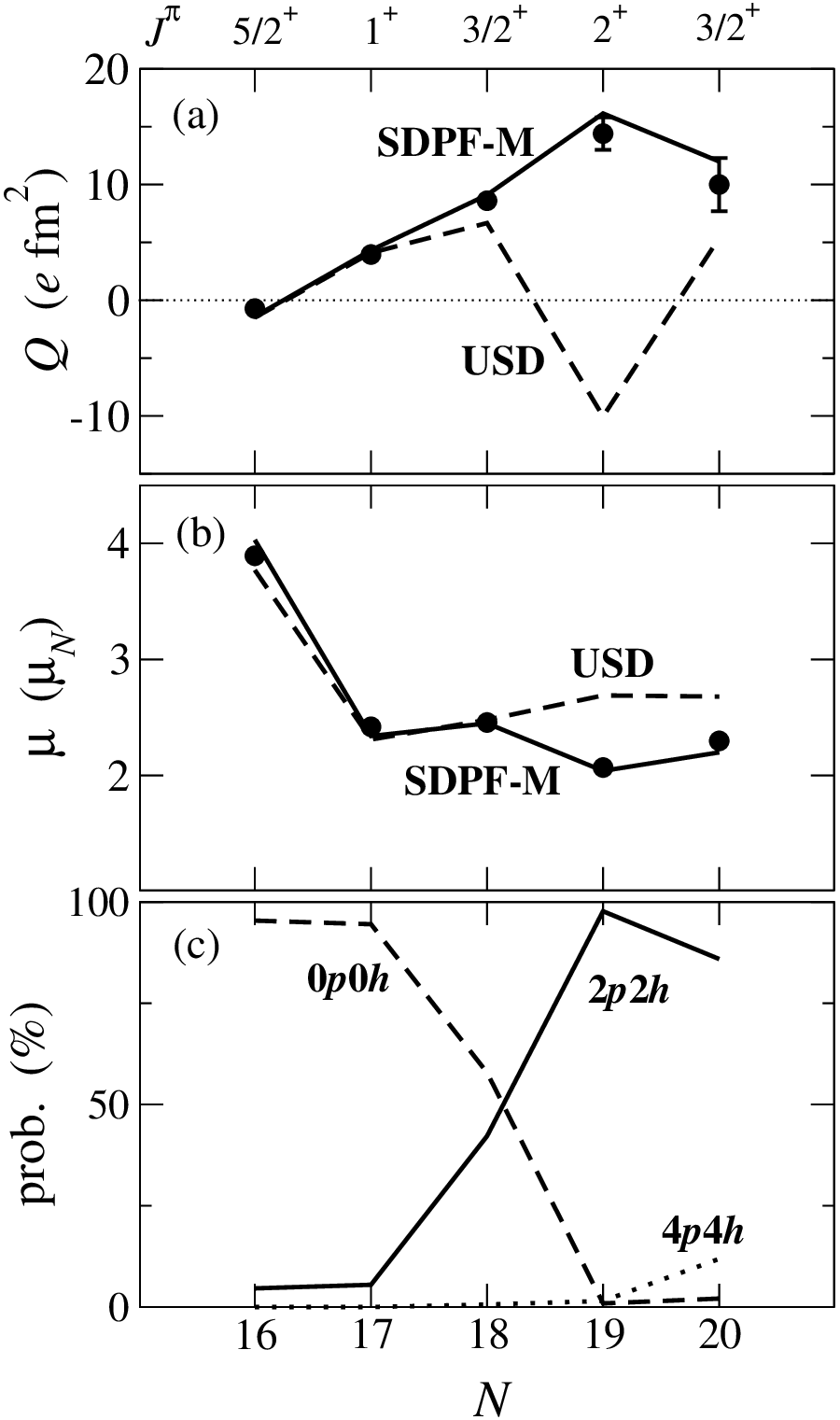}
 \caption{(a) Electric quadrupole moments, (b) magnetic dipole
         moments, and (c) $npnh$ ($n=0,2,4$) probabilities
         of the ground states of neutron-rich Na isotopes,
         as a function of the neutron number, $N$.
         In (a) and (b), the circles are experimental values
         taken from \protect\cite{keim1,keim2},
         while the solid and the dashed lines
         denote, respectively, the MCSM calculation with the SDPF-M
         interaction and USD-model calculation.
         }
          \label{fig:moment}
 \end{center}
\end{figure}

The shell-model calculation described in the above section
is carried out for Na isotopes from $N=16$ to 20.
We first compare, in Fig.~\ref{fig:moment},
the electric quadrupole moments and the magnetic dipole moments
between the MCSM with the SDPF-M interaction and experimental data.
As a reference, results from the USD model are
presented, also.
For the $N=16$ and 17 isotopes, the experimental moments
\cite{namom,keim1,keim2} are well reproduced
by both the shell-model calculations,
reflecting the dominance of the $sd$-shell configurations
in their ground states (see Fig.~\ref{fig:moment} (c)).
It can be inferred, from the agreement with the experimental magnetic
moments, that the present nucleon $g$ factors are reasonable.
At $N=18$, the SDPF-M and USD calculations still give similar magnetic
moments in good agreement with the experiment.
On the other hand, the quadrupole moment
by the SDPF-M is larger by about 30\% than the USD value.
Recently, a very precise measurement of the
quadrupole moments for Na isotopes has been carried out
by Keim {\it et al.}
\cite{keim1,keim2}. The measured quadrupole moment of $^{29}$Na is
8.6(3) $e$\,fm$^2$ in a good agreement with the SDPF-M prediction,
9.1 $e$\,fm$^2$.  On the other hand, the deviation of the USD result
for $^{29}$Na from experiment seems somewhat larger than that for typical
$sd$-shell nuclei \cite{usd}.
The situation is almost unchanged if the radial wave function is
replaced with the Hartree-Fock one
or if the isovector effective charge is tuned \cite{keim2}.
It was thus suspected in \cite{keim2} that the experimental quadrupole
moment of $^{29}$Na might indicate some influence from the
intruder configurations.
The present MCSM calculation indeed shows, in Fig.~\ref{fig:moment} (c),
the large mixing of intruder configurations
by $\sim 42$\%, and their effects
are visible in the quadrupole moment.

Unlike the cases for $N<19$, in the cases of $N=19$ and 20,
the moments
cannot be reproduced by the USD model at all.
Let us start with the most unstable isotope, $^{31}$Na.
The $^{31}$Na nucleus ($N=20$) has been known as a typical case of the
intruder dominance in the ground state \cite{campi}. Its magnetic moment
was reproduced by previous shell-model calculations
in a large shell-model space by Fukunishi {\it et al.}
\cite{fuku} and by Caurier {\it et al.}  \cite{caurier-n20},
supporting this picture.
As shown in Fig.~\ref{fig:moment} (c),
the present calculation, allowing full configurations
within the valence shell, confirms the intruder dominance
in $^{31}$Na and indicates some mixing of even higher intruder
configurations. Accordingly, we can reproduce not only
the magnetic moment
but also the quadrupole moment \cite{keim1}.
Note that the prediction of the energy of
the first excited state of $^{31}$Na \cite{mcsm-f} is
in agreement with the
measurement by intermediate-energy
Coulomb excitation \cite{na31coul}.

We shall now move on to $^{30}$Na ($N=19$), which is the most
crucial nucleus in this paper.
The ground-state property of $^{30}$Na had been rather
obscure so far.
The observed two-neutron separation energy shows no deviation
from the USD-model systematics.
The ground-state spin $J=2$ \cite{namom}
can be explained by the USD model.
This is in contrast to
the anomalous $J=3/2$ ground state in $^{31}$Na,
which is not obtained by the USD model.
The  experimental magnetic moment of $^{30}$Na
2.083(10) $\mu_N$ \cite{namom}, however, deviates from
the USD-model value 2.69 $\mu_N$.
This deviation seems to be somewhat larger than typical deviations in
$sd$-shell nuclei.
As Fig.~\ref{fig:moment} (b) shows, this deviation is
resolved by the MCSM with the SDPF-M interaction
as a consequence of the intruder ground state (see Fig.~\ref{fig:moment} (c)).
As the calculated magnetic moments of $^{30,31}$Na can be changed
only less than by 0.1 $\mu_N$ by replacing the effective
$g$ factors with the free nucleon ones,
the agreement with the experiment
should not be attributed to the choice of the $g$ factors.
Recently, the quadrupole moment has been measured also
by Keim et al. \cite{keim1}. This value, even its sign,
turns out to be quite different from the USD prediction.
Figure~\ref{fig:moment} (a) indicates that the MCSM with the SDPF-M
interaction indeed reproduces this quadrupole moment, too.
Therefore, the properties of the electromagnetic moments
indicate that, in Na isotopes, the ground state
is dominated by the intruder configurations at $N=19$ ($^{30}$Na),
and intruder configurations are substantially mixed
in the ground state already at $N=18$.

\begin{figure}
 \begin{center}
 \includegraphics[width=7.0cm,clip]{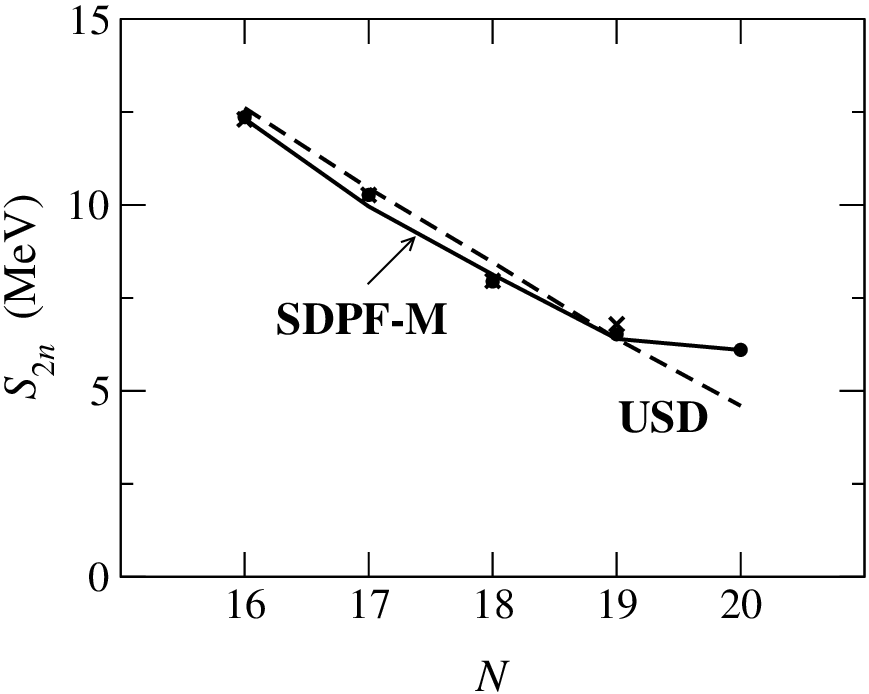}
 \caption{ Two-neutron separation energies of Na isotopes,
  as a function of the neutron number, $N$.
  The circles and the crosses are the experimental values
  taken from the mass table by Audi {\it et al.} \protect\cite{audi}
  and a new measurement by Lunney {\it et al.} \protect\cite{na-mass},
  respectively. The solid line denotes the MCSM calculation with the
  SDPF-M interaction, while the dashed line the USD-model calculation.
         }
          \label{fig:s2n}
 \end{center}
\end{figure}

It may be of interest to discuss the binding energies of Na isotopes
to some detail,
because the USD model explains the observed trend of
binding energies up to $N$=19 rather well.
Figure~\ref{fig:s2n} compares the
two-neutron separation energies ($S_{2n}$) of Na isotopes between the
experiment \cite{audi,na-mass} and the shell-model
calculations.
It can be seen that the USD gives an overall agreement with
experiment as well as the MCSM calculation with the SDPF-M interaction,
except for the failure by USD at $N=20$.
This problem at $N=20$ has been known for many years,
as discussed in Sec.~\ref{sec:intro}. In fact,
as previous models (see, e.g., \cite{campi,island,caurier-n20,mcsm-f})
indicated, the shortage
of the $S_{2n}$ of $^{31}$Na by 1.5~MeV in the USD model
is remedied by having the intruder-dominant ground state.
On the other hand,
the $S_{2n}$ value at $N=19$ can be reproduced well by both the
USD and SDPF-M, whereas their wave functions
are completely different as can be seen in Fig.~\ref{fig:moment} (c).
We shall now resolve this puzzle of $^{30}$Na.

\begin{figure}
 \begin{center}
 \includegraphics[width=7.0cm,clip]{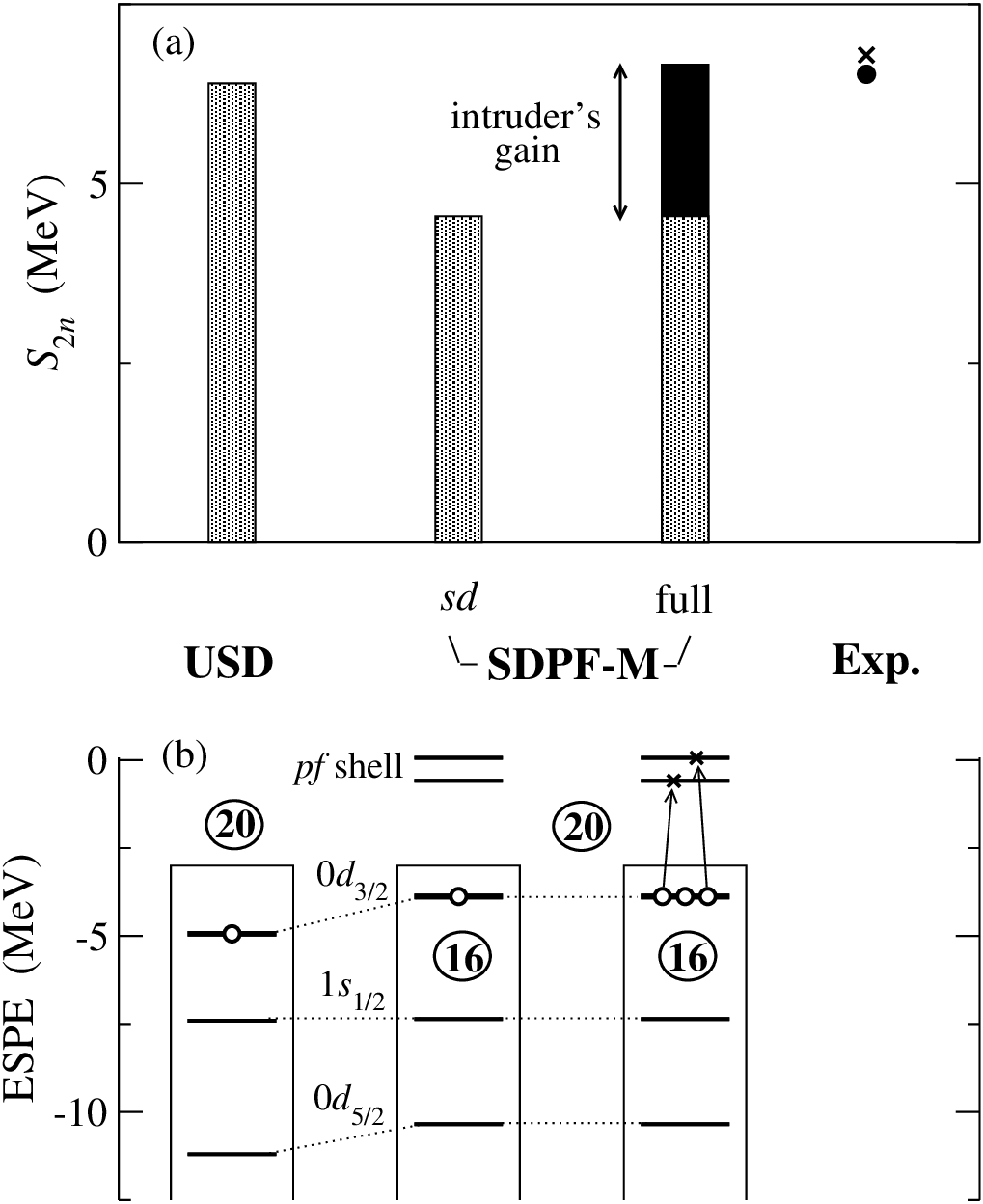}
 \caption{ (a) $S_{2n}$ of $^{30}$Na
  compared among the shell-model calculations
  (with the USD interaction and the SDPF-M one) and experiment.
  For the SDPF-M interaction, a truncated calculation
  within the $sd$ shell
  and the full one by the MCSM are compared, too.
  The circle and the cross are experimental data taken from
  \cite{audi} and \cite{na-mass},
  respectively.
  (b) Corresponding dominant neutron configurations of the ground
   state and the ESPE's
   obtained from each interaction. All the ESPE's are obtained by
   assuming the filling configuration.
         }
          \label{fig:espe}
 \end{center}
\end{figure}

Figure~\ref{fig:espe} (a) compares the experimental $S_{2n}$ of $^{30}$Na
with the calculated values by using the USD interaction and the
SDPF-M interaction.
With the SDPF-M interaction, we carry out two
calculations, i.e., a truncated shell model within the $sd$ shell and
the full calculation.
The results from the USD and the SDPF-M within the $sd$ shell show
rather different $S_{2n}$ values of $^{30}$Na, despite the same model
space.
In order to understand this difference,
the ESPE is considered as shown in Fig.~\ref{fig:espe} (b).
In the SDPF-M interaction, the ESPE of the $0d_{3/2}$ for
small $Z$ is higher than that of the USD interaction.
This difference is a consequence of the shell evolution mentioned
earlier, and is the largest near $Z=8$, creating a new $N=16$ magic number.
The neutron $0d_{3/2}$ orbit is lowered
as $Z$ becomes larger, due to
the strong spin-isospin dependence of the $NN$ interaction \cite{magic}.
At $Z=11$ (Na), this $0d_{3/2}$ is still rather high.
Thus, if the calculation is restricted to the $sd$ shell,
the SDPF-M interaction
produces smaller $S_{2n}$ than that of the USD
for the nuclei where the last neutron is in the $0d_{3/2}$.
On the other hand,
the intruder configurations dominate the
ground state in the full calculation by the MCSM, increasing the
binding energy and
making $S_{2n}$ larger to the same extent as the USD calculation
in the $sd$ shell.
Thus, almost the same separation
energies can be obtained from different mechanisms,
and one has to combine other physical observables
to draw definite conclusions.

\section{Energy levels}\label{sec:level}

\begin{figure*}[t]
 \begin{center}
 \includegraphics[width=13.0cm,clip]{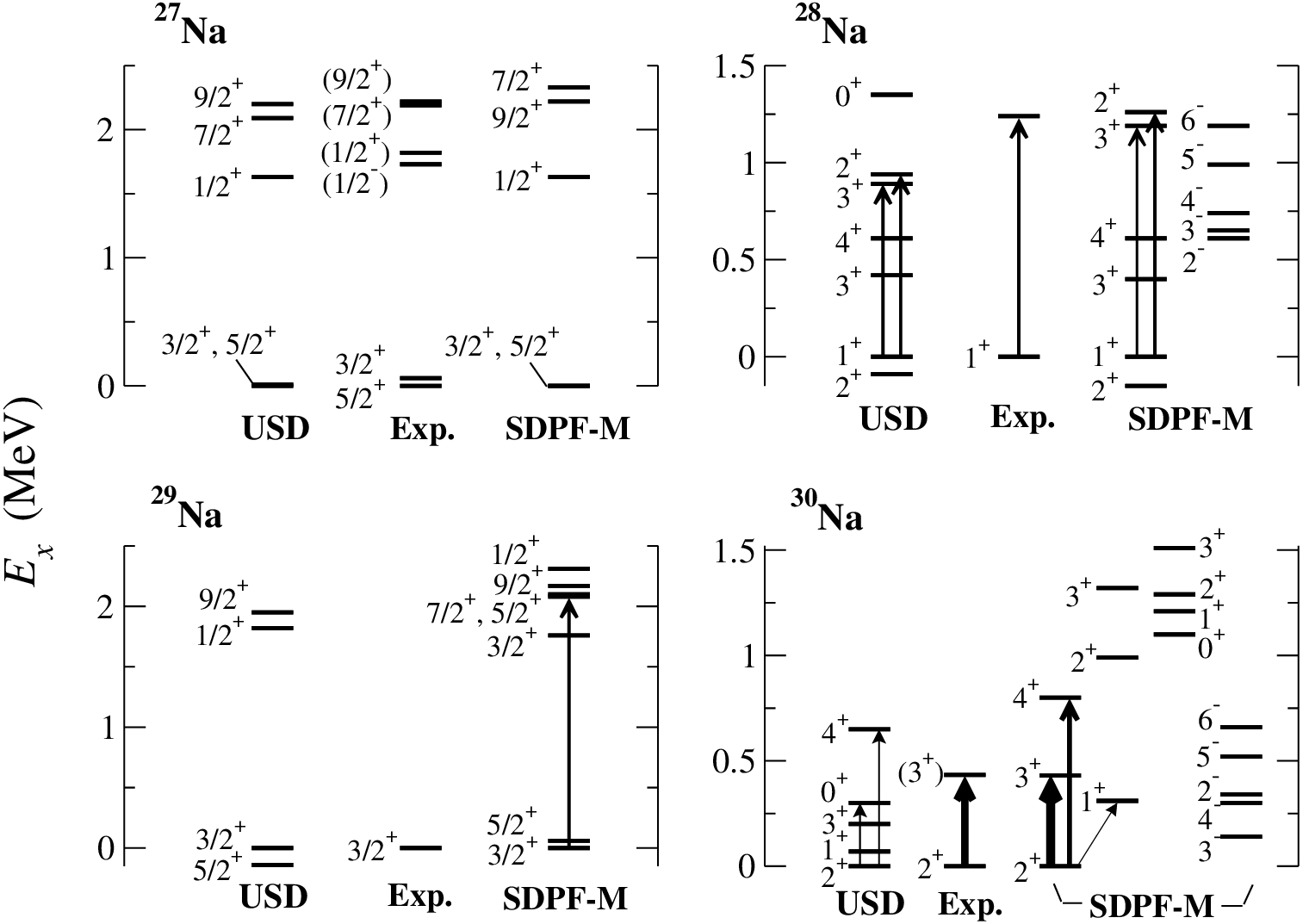}
 \caption{ Comparison of the energy levels of $^{27-30}$Na relative
          to the experimental ground state among
          the experiment (Exp.) and the shell-model calculations
          by the SDPF-M and the USD interactions.
          The $E2$ strength from the ground state
          is illustrated by the width of the arrow. The experimental
          $B(E2)$ values of $^{28,30}$Na and the energy levels of $^{27}$Na
          are taken from \protect\cite{na30coul} and
          \protect\cite{na27level}, respectively.
          For $^{30}$Na, the levels calculated from SDPF-M interaction
          are grouped into four columns; the first (second) one is
          $K$=2 (1) rotational band dominated by intruder configurations,
          the third one represents spherical states which are basically
          of normal configurations, and negative-parity states are shown
          in the fourth column.
         }
          \label{fig:level}
 \end{center}
\end{figure*}

The energy levels of $^{27-30}$Na
are calculated for the SDPF-M interaction by the MCSM, and are
compared with both the experiment and the USD model
in Fig.~\ref{fig:level}. Note that those of $^{31}$Na have been
reported in \cite{mcsm-f}, and are not included here.
Although there have been just few experimental levels
published so far, they provide us with important information.
We shall present, with emphasis on the intruder configurations,
predictions from the SDPF-M interaction, which can be some
help for future experiments.

\subsubsection{$^{27}$Na}
The USD model and the MCSM with SDPF-M give
similar energy levels, and in the latter
a state substantially affected by the intruder
configurations does not appear low.
The calculated energy levels
are in good agreement with the experimental ones
observed recently \cite{na27level} except for the absence of
the 1.725 MeV state (see Fig.~\ref{fig:level}).
The agreement with the experiment confirms high predictive
power of the USD interaction for near-stable nuclei.
The state absent in the calculations has been tentatively
assigned as the $1/2^-$ \cite{na27level}.
As discussed in \cite{na27level}, it is unlikely
that at $N=16$ the negative-parity state dominated by a
one-neutron excitation across the $N=20$ gap appears low,
partly because of a somewhat large gap from the $1s_{1/2}$ to the above
orbits and partly because of the strong pairing correlation in
even-$N$ neutrons.
If this state has a negative parity, it would involve
a one-proton excitation from the $Z=8$ closed shell.

\subsubsection{$^{28}$Na}
The experimental ground state of $^{28}$Na is $J=1$ \cite{namom},
while in the calculations the $1^+_1$ and $2^+_1$ states
are located quite closely in energy but the $2^+_1$ is slightly
lower. The $1^+_1$, $2^+_1$, $3^+_1$ and $4^+_1$ states are
dominated by the configurations consisting of
a neutron $\nu (0d_{3/2})^1$ coupled weakly to the
proton $J=3/2^+$ or $5/2^+$ state (see the energy levels of $^{27}$Na),
and are close to one another.
For these states, both the shell-model calculations
give similar excitation energies.

A recent Coulomb-excitation experiment
by Pritychenko {\it et al.} \cite{na30coul}
shows a $\gamma$ ray at 1.24 MeV with
$B(E2)\uparrow = 54(26)$ $e^2$\,fm$^4$.
The MCSM with SDPF-M gives $B(E2; 1^+_1 \to 2^+_2)= 69$
$e^2$\,fm$^4$ and $B(E2; 1^+_1 \to 3^+_2)= 47$ $e^2$\,fm$^4$,
either (or the sum)
of which may correspond to the observed $\gamma$ ray.
On the other hand, the $B(E2; 1^+_1 \to 2^+_1)$  and
$B(E2; 1^+_1 \to 3^+_1)$ are as small as
19 and 27 $e^2$\,fm$^4$, respectively.
Similar $B(E2)$ values are obtained by the USD model,
but the relevant $2^+_2$ and $3^+_2$ energy levels by the USD
are lower by $\sim 0.3$ MeV than those of the SDPF-M
(see Fig.~\ref{fig:level}).
By analyzing the occupation numbers of the wave functions,
it turns out that the $2^+_2$ and $3^+_2$ states
are mainly composed of one-neutron excitation from the $1s_{1/2}$
to the $0d_{3/2}$.
As the gap between these orbits is larger for the
SDPF-M interaction, those states are pushed up.

The negative-parity states are predicted to
lie rather low reflecting the narrower $N=20$ shell gap,
but there is no experimental information presently.
They might be compared qualitatively to
the state at 1.095 MeV in the $N=17$ isotone $^{29}$Mg
which can be a negative-parity state
as discussed by Baumann {\it et al.} \cite{mg29}.

\subsubsection{$^{29}$Na}
The ground state of $^{29}$Na is $J=3/2$ experimentally
\cite{namom}. The calculations show very close $3/2^+_1$ and
$5/2^+_1$ levels, and the MCSM gives the correct spin order,
whereas the USD model does not (see Fig.~\ref{fig:level}).
This difference is because the $3/2^+_1$ state contains a
larger mixing of the intruder configurations than the $5/2^+_1$.
The shell-model calculations show that
the $5/2^+_1$ state is strongly connected to the ground state
with $B(E2; 3/2^+_1 \to 5/2^+_1)=111$ $e^2$\,fm$^4$  by the
USD model (135 $e^2$\,fm$^4$ by the MCSM),
while the $B(E2)$ values from the ground state to
the other normal-dominant low-lying states are very small.
We thus point out that the Coulomb-excitation would hardly
populate other excited states as far as the low-lying states
are dominated by normal configurations.

In the USD model, it is predicted that there are just $1/2^+_1$ and
$9/2^+_1$ levels around 2 MeV.
Apart from these normal-dominant states,
the MCSM predicts, around the same energy, low-lying $3/2^+_2$, $5/2^+_2$ and
$7/2^+_1$ states dominated by the intruder configurations.
Due to the large mixing in the ground state,
the $7/2^+_1$ may be excited by the Coulomb excitation
with a moderately large value,
$B(E2; 3/2^+_1 \to 7/2^+_1)=57$ $e^2$\,fm$^4$, as
predicted by the MCSM.

\subsubsection{$^{30}$Na}
Both the calculations succeed in reproducing the
ground-state spin, but the energy levels are quite
different. In the USD model, the low-lying states
are composed mainly of the configurations with a neutron hole
$\nu (0d_{3/2})^{-1}$ coupled weakly to the proton
$J=3/2$ or 5/2 state.
The $E2$ strength between them
should then be weak as depicted in Fig.~\ref{fig:level}.
On the other hand, the MCSM with SDPF-M gives the intruder-dominant
ground state which is strongly deformed.
Indeed, we obtain a rotational band connected by
strong $E2$ transitions: the $E2$ matrix elements calculated
by the MCSM linked to the ground state are
$B(E2; 2^+_1 \to 3^+_1)= 168$ $e^2$\,fm$^4$,
$B(E2; 2^+_1 \to 4^+_1)= 90$ $e^2$\,fm$^4$, and
$Q(2^+_1)= 16$ $e$\,fm$^2$.
They give rise to similar intrinsic quadrupole moments,
i.e., $Q_0= 58$, 65 and 56 $e$\,fm$^2$, respectively,
by assuming $K=2$.
The strong $E2$ transition has recently been measured
by the Coulomb-excitation experiment
by Pritychenko {\it et al.} \cite{na30coul}: from
the strength of the measured $\gamma$ ray the $B(E2\uparrow)$
was deduced to be $130^{+90}_{-65}$ $e^2$\,fm$^4$ consistently with
the MCSM calculation.
The anomalous quadrupole moment (see Fig.~\ref{fig:moment})
and this large $B(E2)$ value in $^{30}$Na are excellently accounted
for as a result of the large prolate deformation
associated with the intruder-dominant configurations.

From the viewpoint of the particle-rotor picture,
the intrinsic state of the yrast band is regarded as
a proton in the $\pi [211]3/2^+$ Nilsson orbit and a neutron
in the $\nu [200]1/2^+$ coupled to a
deformed $^{28}$Ne rotor. As a result, $K=1$ and 2 are
possible as the yrast band, and the MCSM shows that
the latter is favored in energy.
It is of interest to point out that this feature is consistent
with the so-called Gallagher-Moszkowski rule \cite{gallagher}
that in strongly deformed nuclei
the favored $K$ is made so that
the intrinsic spins of the last proton
and neutron are parallel.
Thus, the agreement of the ground-state spin $J=2$ by the MCSM
is not just an accidental fortune reflecting a particular interaction
matrix element, but has been conducted
once the intruder configurations dominate the ground state.

The MCSM yields also the $K=1$ band starting at 0.31 MeV.
Its $J=2$ and 3 members are calculated to lie
around 1 MeV as shown in Fig.~\ref{fig:level}, while they are
well mixed with the normal-dominant states.
Also at 1 $\sim$ 1.5 MeV excitation energy,
normal-dominant spherical states, corresponding to the lowest
states in the USD model, appear as shown in Fig.~\ref{fig:level}.
The negative-parity states are predicted to be rather low,
dominated by the $1p1h$ excitation across the $N=20$ shell gap.
The competition between normal and intruder configurations in $^{30}$Na
seems to be very intriguing, and is discussed in
the next section in more detail.

\section{Shell-gap dependence on the intruder dominance}\label{sec:gap}

From the above discussions on the moments and the levels,
it becomes evident that the transition from the
normal to intruder ground state occurs fully at
$N=19$, after strong normal-intruder mixing already at $N=18$.
We shall show, in this section, that this normal-intruder transition
for $N <$ 20 is particularly sensitive to the shell gap.

In general, an intruder state can be the ground state,
if the energy gain due to dynamical correlations including
deformation overcomes the energy loss in transcending nucleons
across the shell gap.
The shell gap is nothing but the difference between ESPE's of relevant orbits.
The neutron ESPE changes rather gradually as a function of
the neutron number, since the monopole interaction
for $T=1$ is weak. Namely, the neutron shell gap is rather constant
as a function of the neutron number.
This implies that what is crucial
in the transition from a normal to an
intruder ground state within an isotope chain is primarily
the neutron-number dependence of the correlation energy and
its relative magnitude to the shell gap.
Here, a good index of the correlation energy is
the difference between the eigenvalue of the total Hamiltonian
and the expectation value of the monopole interaction for
the filling configuration.

\begin{figure}[t]
 \begin{center}
 \includegraphics[width=8.0cm,clip]{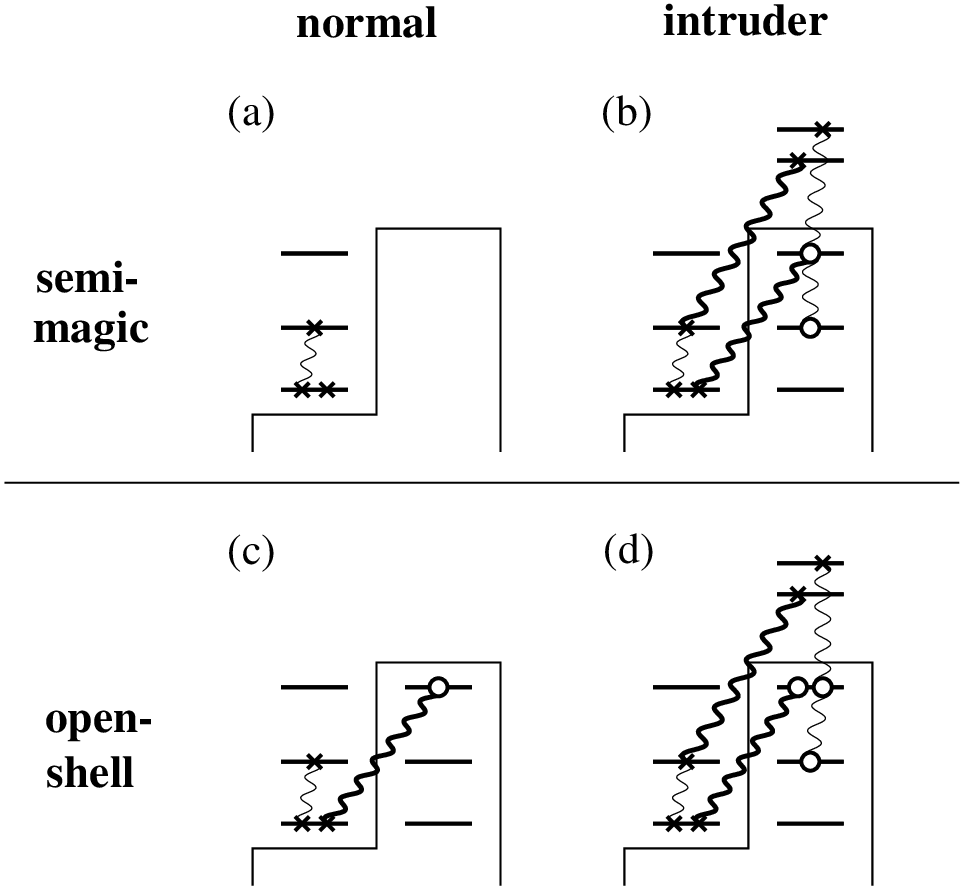}
 \caption{ Schematic sketch of the sources of the correlation energy
  of the intruder and the normal states
  of semi-magic ((a) and (b)) and open-shell ((c) and (d)) nuclei.
  Typical configurations for these states are shown.
  The proton-neutron interaction is illustrated by thick wavy lines,
  while the proton-proton and neutron-neutron interactions are
  drawn by thin wavy lines.
         }
          \label{fig:corr}
 \end{center}
\end{figure}

In Fig.~\ref{fig:corr}, the sources of the correlation energy
are sketched schematically.
Since a normal state of a (neutron) semi-magic nucleus
consists of configurations shown in Fig.~\ref{fig:corr} (a),
only the proton rearrangement is relevant to
the correlation energy, which is generally small.
On the other hand, the correlation energy is very large
in the case of an intruder state
composed of configurations like Fig.~\ref{fig:corr} (b),
due to large numbers of particles and holes in active orbits.
We note that the proton-neutron interaction produces much larger
correlation energies than the interactions between like nucleons.
This makes the correlation energy in Fig.~\ref{fig:corr} (b) much
larger than that of Fig.~\ref{fig:corr} (a), favoring
the normal-intruder inversion even with a large shell gap.

On the other hand, in the cases like Fig.~\ref{fig:corr} (c) and (d),
a normal state of an open-shell nucleus has a neutron hole already.
The neutron rearrangement is then possible, and strong
proton-neutron two-body matrix elements contribute to
the correlation energy.
The intruder configurations of Fig.~\ref{fig:corr} (d)
gain correlation energy similarly to the case of Fig.\ref{fig:corr} (b).
However, the difference of the correlation energy between
Fig.~\ref{fig:corr} (a) and (b) is larger than that between
Fig.~\ref{fig:corr} (c) and (d), because of the saturation
of the correlation energy with many particles and many holes as
is the case in Fig.~\ref{fig:corr} (d).
A concrete example can be found with the USD interaction:
a semi-magic $^{31}$Na gains the correlation energy only by
1.7 MeV within the $sd$ shell, whereas it increases to
3.7 MeV for $^{30}$Na and further to 7.2 MeV for $^{29}$Na.
The correlation energy of intruder states increases more slowly
due to the saturation as mentioned just above.
This implies that the intruder dominance in $N<20$ nuclei
becomes less favored as $N$ goes down from 20.  Hence,
if the normal-intruder inversion still occurs, it should be due to
a narrower shell gap.  We shall present a more detailed account on
this point now.

The SDPF-M interaction indeed gives
a narrow $N=20$ shell gap ($\sim$ 3 MeV) for Na isotopes.
Note that it still reproduces
the large gap ($\sim$ 6 MeV) of $^{40}$Ca, owing to its monopole property.
We now demonstrate how such a narrow gap of Na isotopes plays
a crucial role in the intruder dominance in $^{30}$Na,
by means of a simulation based on the argument just above.
Namely, we vary the shell gap from the value given by the SDPF-M interaction
to larger values, to see what happens.
This can be done by changing the monopole interaction between
the $0d_{5/2}$ and the $0d_{3/2}$ as
\begin{equation}\label{eq:x}
 \delta V_{0d_{5/2},0d_{3/2}}^{T=1,0}(x) = -0.3x, +0.7x\,\, {\rm MeV},
\end{equation}
where $V_{ij}^{T}$ denotes the monopole interaction
between $i$ and $j$ orbits with isospin coupled to $T$ \cite{mcsm-n20}.
The parameter $x$ is to control the ESPE: $x=0$ represents the
situation with the SDPF-M interaction as a starting point.
A larger $x$ means primarily a lower neutron $0d_{3/2}$ level, i.e.,
a wider $N=20$ gap in $^{30}$Na.
Other effects are minor in this nucleus.

\begin{figure}
 \begin{center}
 \includegraphics[width=7.0cm,clip]{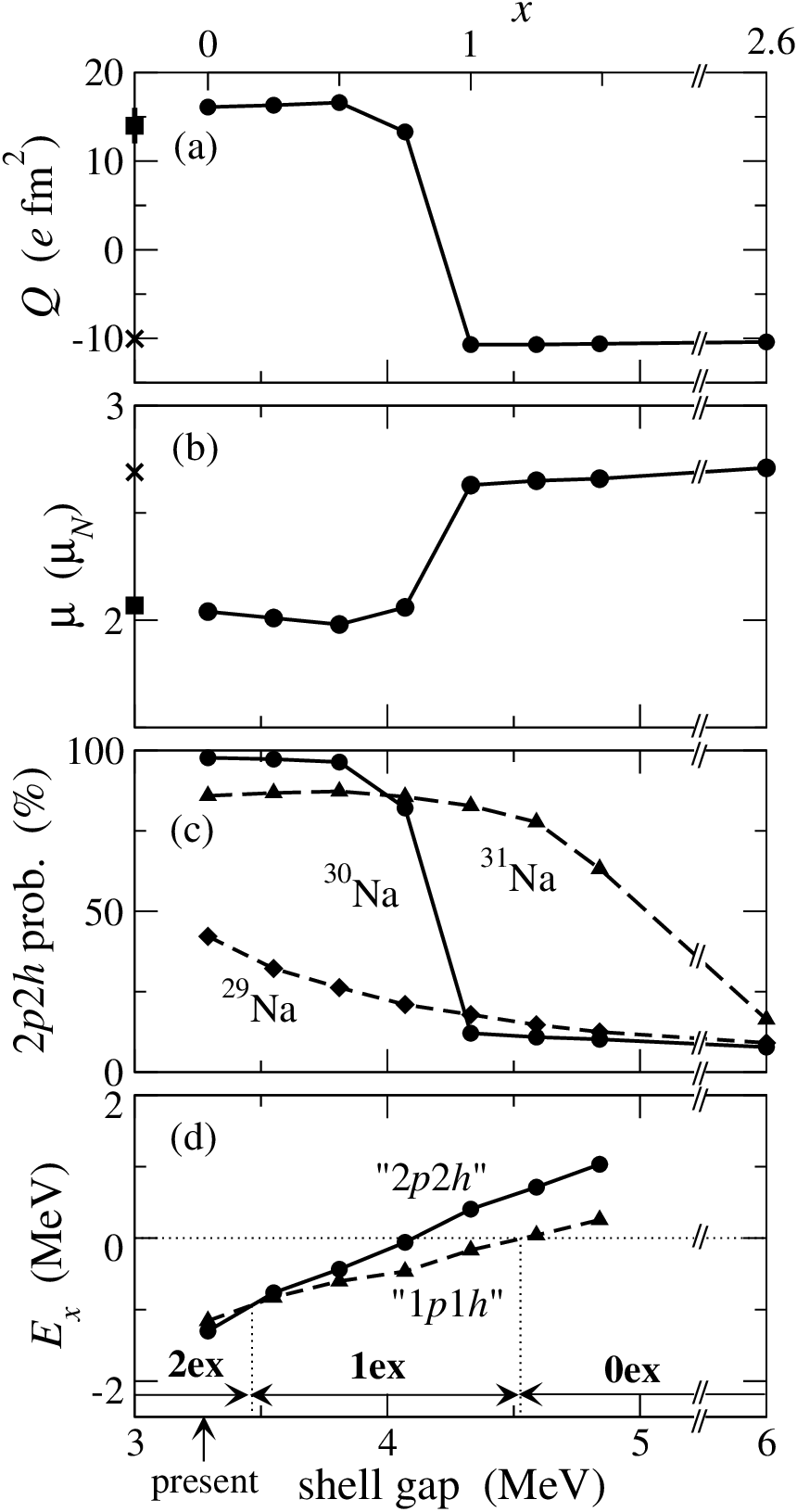}
 \caption{(a) Quadrupole moment and (b) magnetic moment of $^{30}$Na
  as a function of the $N=20$ shell gap
  (controlled by a parameter $x$ in Eq.~(\ref{eq:x})).
  The experimental \protect\cite{keim1,keim2} and USD-model
  ones are denoted by the squares and
  crosses, respectively.
  (c) $2p2h$ probabilities in the (positive-parity)
  ground states of $^{29-31}$Na.
  (d) Energies of the $2p2h$- and $1p1h$-dominant lowest states
   of $^{30}$Na
   (denoted by $''2p2h''$ and $''1p1h''$, respectively)
   measured from that of the $0p0h$-dominant lowest state.
   The range of the shell gap giving the $npnh$ ground state
   ($n=0,1,2$) is indicated by $n$\,ex. Note that the corresponding
   shell gaps of $^{29,31}$Na are, respectively, smaller and larger
   by 0.24~MeV than the one of $^{30}$Na.
       }
       \label{fig:varspe}
 \end{center}
\end{figure}

Figure~\ref{fig:varspe} presents the variation of the
ground-state properties of Na isotopes
as a function of the gap thus varied. Results are mainly about $^{30}$Na
unless otherwise specified.
The SDPF-M interaction (at $x=0$)
gives a narrow $N=20$ shell gap, i.e., 3.3 MeV for $^{30}$Na.
As $x$ is increased, the gap becomes wider, and the
$^{40}$Ca gap ($\sim 6$ MeV) is given by $x\sim 2.6$.  An
intermediate value $x=1$ reproduces the gap of USD ($\sim 4.3$ MeV),
implying that the USD includes some fractional effects of the current
shell evolution \cite{magic}.

The quadrupole moment and the magnetic moment in
Fig.~\ref{fig:varspe} (a) and (b) are almost constant
up to the~gap~$\sim$~4~MeV, as $x$ is increased from 0.
But, it jumps to  values comparable to that of the USD model around
the 4 MeV gap.
In order to see how this rapid transition occurs,
Fig.~\ref{fig:varspe} (c) shows the probability of the
intruder configurations in the lowest positive-parity state for
$^{30}$Na.
As expected from the change of the moments,
the dominant component of the ground state
moves rapidly from intruder to normal configurations
at the shell gap $\sim$~4~MeV.

It is interesting to compare this transition in $^{30}$Na
with the ones of $^{29,31}$Na shown in Fig.~\ref{fig:varspe} (c).
Compared to the pattern of $^{30}$Na, notable differences are that
(i) the shell gap causing the transition is larger
in $^{31}$Na ($\sim 5$~MeV) and smaller in $^{29}$Na
($\sim 3$~MeV), and (ii) the transition takes place more slowly
than $^{30}$Na.
The former is because of the difference of the correlation
energies in the normal states of $^{29-31}$Na discussed already,
and the latter is because
even-$N$ configurations are strongly connected with the
pair-excited states via the pairing interaction.
This is the reason why the intruder dominance in $^{30}$Na
has a particular importance to clarify the shell structure
of Na isotopes, and we now confirm that the narrower shell gap
due to the shell evolution \cite{magic} plays a crucial role.
Note that at the gap of stable nuclei ($\sim 6$~MeV)
the intruder dominance does not occur even in $^{31}$Na.

We finally discuss the competition of the dominant configurations
in $^{30}$Na including a negative-parity state.
Figure~\ref{fig:varspe} (d) displays the energies of
the $2p2h$- and $1p1h$-dominant lowest states
measured from the energy of the $0p0h$-dominant state,
as the gap is changed.  In Fig.~\ref{fig:level}, we can see what
happens as $x$ is increased from 0.  The $2^+_1$ state
is close to the $3^-_1$ but stays lower consistently with
experiment.
If the gap is made larger, the ground state is switched to
a negative-parity state around at 3.5~MeV, and
persists for a while.
At a larger shell gap $\sim 4.5$~MeV,
a competition between a positive-parity state
and a negative-parity one is encountered again,
where the former is dominated by
normal configurations.
Finally, after this competition
the normal-dominant ground state persists.
The ``island of inversion'' picture \cite{island}
seems to correspond to the gap near
the second competition (i.e., around 4.5 to 5~MeV):
with the weak-coupling approximation, the $1\hbar\omega$
and $2\hbar\omega$ states were calculated to be located
at 0.306 and 0.776 MeV above the normal one,
respectively \cite{island}.

The competition of normal- and abnormal-parity states in
$^{30}$Na ($N=19$)
can be compared to a famous example of the parity inversion
in $^{11}$Be ($N=7$). Both are related to the narrow shell gap,
but we point out a large difference between them:
the latter is considered as
the competition between the $0p0h$ and $1p1h$
states, corresponding to the second competition in the present paper.
Thus, in the case of the $N=20$ region a more drastic
event occurs in spite of the normal-parity ground state
of $^{30}$Na,
reflecting a further narrowing of the shell gap.

\section{summary}\label{sec:summary}

In summary, we have investigated where the disappearance of
the magic structure starts in the isotope chain of Na
referring to its mechanism.
It is suggested that
experimental electromagnetic moments, energy levels, and
$B(E2)$ values of $^{30}$Na with $N=19$ clearly indicate the dominance of
the intruder configurations in its ground state,
by combining with a shell-model calculation
using the MCSM. The present result is in sharp contrast to
a previous speculation based on the USD model \cite{usd}
from the viewpoint of the binding energy,
where the disappearance was supposed to occur
right at $N=20$. The same conclusion as this speculation
was drawn by the ``island of inversion'' model.
The difference between the present calculation and the previous
models is mainly in the behavior of the effective $N=20$ shell gap
for small $Z$:  the gap is substantially narrow (about 3 MeV)
for Na isotopes with the present SDPF-M interaction.  Nevertheless,
owing to the monopole part of the SDPF-M interaction, in particular, its
spin-isospin dependent component, the well-known 6 MeV gap is restored
for $^{40}$Ca, as an example of the shell evolution
in stable and unstable nuclei \cite{magic}.
The validity of this argument has been confirmed in this study
quite transparently with Na isotopes below $N=20$ 
where the intruder states
are shown to be, most likely, unable to beat the normal states without
a narrower shell gap.

\acknowledgements

This work was supported mainly by Grant-in-Aid for Specially
Promoted Research (13002001) from the MEXT, by
the RIKEN-CNS collaboration project
on large-scale nuclear structure calculation, and by the U.S. National 
Science Foundation under Grant No. PHY-0110253.
One of the authors (Y.U.) is grateful to Dr. T.~Shizuma for
discussions about the collective property of odd-odd nuclei.
He acknowledges the support in part by
Grant-in-Aid for Young Scientists (14740176) from the MEXT,
and the Helios computer system in JAERI.
The conventional shell-model calculation was carried out by
the code {\sc oxbash} \cite{oxbash}.

\end{document}